\documentstyle[11pt,epsf]{article}

\advance \topmargin by -\headheight
\advance \topmargin by -\headsep
\evensidemargin \oddsidemargin
\begin{document}

\begin{titlepage}
\begin{center}
\section*{The Coulomb Interaction between
 Pion-Wavepackets:\\ The $\pi^+$--$\pi^-$ Puzzle}

\vspace{0.5cm}

H. Merlitz$^{a,b}$ and D. Pelte$^{a,c}$\\
\end{center}

\vspace{0.5cm}

  $a$: Physikalisches Institut der Universit\"at Heidelberg,
  D-69120 Heidelberg\\

  $b$: Gesellschaft f\"ur Schwerionenforschung (GSI) Darmstadt, Germany\\

  $c$: Max-Planck-Institut f\"ur Kernphysik, D-69117 Heidelberg\\

\begin{abstract} 
 The time dependent Schr\"odinger equation for $\pi^+$--$\pi^-$
 pairs, which are emitted from the interaction zone in 
 relativistic nuclear collisions, is solved using wavepacket states.
 It is shown that the Coulomb enhancement in the momentum correlation
 function of such pairs is smaller than obtained in earlier 
 calculations based on Coulomb distorted plane waves. These results 
 suggest that the experimentally observed positive correlation signal
 cannot be caused by the Coulomb interaction between pions emitted
 from the interaction zone. But other processes which involve 
 long-lived resonances and the related extended source dimensions
 could provide a possible explanation for the observed signal.\\
 PACS: 25.70.-z\\
\end{abstract}
\end{titlepage}

\section{Introduction}
 \label{intro}

 Originally, the method of intensity-interferometry (``Hanbury-Brown and
 Twiss interferometry'', HBT) was developed in astronomy and applied
 to determine the angular radii of stars, which were
 too small to be resolved in ordinary optical instruments \cite{hbt57}.
 Shortly afterwards it was recognized that a similar procedure
 could also be applied to pions emitted in high energy reactions,
 whereby the size of the pion-production volume becomes a measurable
 quantity \cite{gol60}. However, it is clear that the different physical
 environment in high energy- or heavy ion reactions
 asks for certain modifications of the original formalism. The following 
 facts have to be taken into account:
 \begin{enumerate}
  \item The wavelength $\lambda$ of visible light is about $500$ nm and
    the coherence length of a photon
    $\Lambda \approx \tau \, c$ with a typical emission
    time of $\tau \approx 10^{-8}$ s is about $7$ orders of magnitude larger.
    Therefore, the photon can be  well described as a plane wave.
    A typical pion, produced via $\Delta$(1232)-decay 
    with a momentum $p_{\rm o} = 150$ MeV/c has the de Broglie 
    wavelength $\lambda_{\rm o} \approx 8$ fm, and the coherence length
    is defined as (\cite{hol95}, p.\ 158) 
    $\Lambda = \lambda_{\rm o}^2/\Delta \lambda \approx 6$ fm. 
    Here, $\Delta \lambda$ 
    is the uncertainty in the wavelength, related to the energy width
    $\Delta E \approx 120$ MeV of the $\Delta$(1232) 
    resonance via $\Delta \lambda
    = 2\pi\,\hbar/\Delta E$. Since $\lambda_{\rm o}$ and $\Lambda$ are of 
    the same size, the pion cannot be described as a plane wave.
 \item The radius $R$ of a star is typically $8$ orders of magnitude larger than
    the coherence length of the emitted light, and therefore coherence
    effects at emission time can be neglected: The star is a chaotic
    radiator. The pion source has a typical size of a few fm which is 
    of the same order as the coherence length of the pion. 
    Therefore, interference
    effects in the source have to be taken into account, as it is
    consistently done in the wavepacket model [4-6].
 \item Photons are neutral, whereas the pion correlations 
    (with exception of $\pi^{\rm o}$) are modified by the Coulomb interaction.
 \end{enumerate} 
 The Coulomb distortion of the correlation function and the corresponding
 correction has been the topic of many recent publications [7-10],  
 where the pions where described either as plane waves or as classical 
 particles. The several existing theoretical approaches all have in
 common that they do not consider the zero point energy which emerges
 as a consequence of the finite size of the pion source. However,
 in a recent wavepacket calculation of identical and Coulomb 
 interacting pions 
 it was found that the correlation function does not exhibit any significant
 distortion, a fact which was explained by the strong momentum uncertainty
 of the wavepackets \cite{mer97}. This argument, if valid, should also
 apply to the correlations of $\pi^+$--$\pi^-$ pairs. Here, positive
 correlations were found experimentally \cite{bar97}, which are 
 interpreted as caused by the Coulomb attraction between both pions and
 equivalently used to Coulomb-correct the correlation functions
 for like-charge pions. 

 It is therefore of crucial importance to verify the conclusions of
 Ref.\ \cite{mer97} also for the $\pi^+$--$\pi^-$ system of non-identical
 mesons, where the wavefunction does not have to be symmetric with respect
 to particle exchange. This system therefore displays the consequences
 of the quantum behaviour, and in particular those of the Heisenberg 
 uncertainty principle, most purely, and it allows to study the
 modifications when the system approaches the classical limit.
 To this end, we compute the correlation function of $\pi^+$--$\pi^-$ pairs
 using wavepackets for the pion-states. We again discuss the important role 
 of the dispersion which determines
 the size of the Coulomb distortion. We compare
 the results with experimental data and finally discuss the implications.

\section{The multiconfigurational procedure}
 \label{multi}
 The method to solve the time dependent Schr\"odinger equation
 for $\pi^+$--$\pi^-$ pairs,
 \begin{equation}
  \label{eq10}
  \hat{H} \Psi({\bf r}_1, {\bf r}_2,t) =
  i\hbar \partial_t \Psi({\bf r}_1, {\bf r}_2,t)\;,
 \end{equation}
 with the Hamiltonian
 \begin{equation}
  \label{eq20}
   \hat{H} = -\frac{\hbar^2}{2m}(\Delta_1 + \Delta_2) + \frac{-e^2}{
   |{\bf r}_1 - {\bf r}_2|} 
 \end{equation}
 is similar to the method used in \cite{mer97}. For the 
 initial $2$-particle wavefunction we use the product state
 \begin{equation}
  \label{eq30}
  \Psi({\bf r}_1, {\bf r}_2,t=0) =
  \psi_1({\bf r}_1,t=0)\cdot \psi_2({\bf r}_2,t=0)\;,
 \end{equation}
 where the single particle states are Gaussians, i.\,e.\
  \begin{equation}
   \label{eq40}
   \psi_j({\bf r}_j,t=0) = (2\pi\sigma_{\rm o}^2)^{-3/4}\,
   \exp \left(\frac{i}{\hbar}{\bf P}_j\cdot{\bf r}_j
    -\frac{\left({\bf r}_j - {\bf R}_j\right)^2}
        {4\sigma_{\rm o}^2}\right)
 \end{equation}
 with $j \in \{1,2\}$. ${\bf R}_j$ is the initial position of
 the $j$'th wavepacket-centre, ${\bf P}_j$ the initial momentum  
 and $\sigma_{\rm o}$ the initial 
 wavepacket-width. Transforming into center of mass (com) and
 relative (rel) coordinates, the state is rewritten as
 \begin{equation} 
  \label{eq50}
  \Psi({\bf r}_{\rm c}, {\bf r},t=0) = \Psi_{\rm com}
  ({\bf r}_{\rm c})\cdot \Psi_{\rm rel}
  ({\bf r})\,
 \end{equation}
 with ${\bf r}_{\rm c} = 
 ({\bf r}_1 + {\bf r}_2)/2$ and ${\bf r} = {\bf r}_1 - {\bf
 r}_2$,
 \begin{equation}
   \label{eq60}
   \Psi_{\rm com}({\bf r}_{\rm c}) = (\pi\sigma_{\rm o}^2)^{-3/4}\,
   \exp\left(\frac{i}{\hbar} {\bf P}_{\rm c}\cdot{\bf r}_{\rm c}
   -\frac{\left({\bf r}_{\rm c} - 
    {\bf R}_{\rm c}\right)^2}
        {2\sigma_{\rm o}^2}\right)\;,
 \end{equation}
 and
 \begin{equation}
   \label{eq70}
   \Psi_{\rm rel}({\bf r}) = (4\pi\sigma_{\rm o}^2)^{-3/4}\,
   \exp\left(\frac{i}{\hbar}{\bf P}\cdot{\bf r}
    -\frac{\left({\bf r} - {\bf R}\right)^2}
        {8\sigma_{\rm o}^2}\right)\;.
 \end{equation}
 Here, ${\bf R}_{\rm c} = ({\bf R}_1 + {\bf R}_2)/2$, ${\bf R} = 
 {\bf R}_1 - {\bf R}_2$, ${\bf P}_{\rm c} = {\bf P}_1 + {\bf P}_2$
 and ${\bf P} = ({\bf P}_1 - {\bf P}_2)/2$.
 Since there are no external forces, the time dependent solution
 for $\Psi_{\rm com}$ is the solution for a free wavepacket, i.\,e.\
 \begin{equation}
   \label{eq80}
   \Psi_{\rm com}({\bf r}_{\rm c},t) = (\pi s_{\rm c}(t)^2)^{-3/4}\,
   \exp\left(\frac{i}{\hbar}\left({\bf P}_{\rm c}\cdot{\bf r}_{\rm c}
   - E_{\rm c} t\right)
    -\frac{\left({\bf r}_{\rm c} - {\bf R}_{\rm c}\right)^2}
        {2 s_{\rm c}(t)\sigma_{\rm o}}\right)\;.
 \end{equation}
 Here, $E_{\rm c} = P_{\rm c}^2/(2M)$ with $M = 2m$ and the pion mass
 $m = 140$ MeV, finally
 \begin{equation}
   \label{eq90}
   s_{\rm c}(t) = \sigma_{\rm o}\left(1 + \frac{i\hbar t}
   {M\sigma_{\rm o}^2}\right)\;.
 \end{equation}
 In order to evaluate the time dependent solution for
 $\Psi_{\rm rel}({\bf r},t)$, an expansion into a set of basis
 functions is required. As basic components for our basis we use
 the same time dependent wavefunctions as in \cite{mer97}, i.\,e.\
  \begin{eqnarray}
   &&\psi_{l,m,n}({\bf r},t) = \nonumber \\ 
   &&\sqrt{{\cal N}_{l,m,n}}\,
   x^{l}\, y^{m}\, z^{n}
   \exp\left(
   \frac{i}{\hbar} \left({\bf P}(t)\cdot
   {\bf r} - \theta_{l,m,n}(t) \right) -
   \frac{\left({\bf r} - {\bf R}(t)\right)^2}
        {8s(t)\sigma_{\rm o}}\right)
   \label{eq100}
 \end{eqnarray}
 with the phase
 \begin{equation}
  \label{eq110}
  \theta_{l,m,n}(t) = \int_{0}^{t}E (\tau) \, d\tau +
  \hbar\, (l + m + n + 3/2)\,
  \arctan\left(\frac{\hbar t}{2\mu\sigma_o^2}\right)
 \end{equation}
 and $x^{l} = ({\bf r}_x - {\bf R}_{x})^{l}$, and similar
 for the $y$- and $z$ coordinates. Further, ${\bf P} =
 ({\bf P}_1 - {\bf P}_2)/2$, $\mu$ is the reduced mass, 
 $l$, $m$ and $n$
 are positive integers, 
 \begin{equation}
   \label{eq120}
   s(t) = \sigma_{\rm o}\left(1 + \frac{i\hbar t}
   {4\mu\sigma_{\rm o}^2}\right)
 \end{equation}
 and the norm is
 \begin{equation}
  \label{eq130}
  {\cal N}_{l,m,n} = \frac{(4\pi \sigma^2)^{-3/2}\,
                     (\sqrt{2}\,\sigma)^{-2(l + m + n)}}
                     {(2l - 1)!!\, (2m -1)!!\, (2n -1)!!}
 \end{equation}
 with $\sigma = |s|$ and $j!! = j\, (j-2)\, (j-4)\cdots$.
 We may call a state with 
 $k = l + m + n = 0$ ``s-state'', similar ``p-state''
 for $k = 1$, ``d-state'' for $k = 2$ and ``f-state'' for $k = 3$.
 The basis functions $\psi_{l,m,n}({\bf r},t)$ are not orthogonal
 for different sets $\{l,m,n\}$, but linear combinations
 of them can be found which are orthogonal \cite{cle89}.
 We use the following orthogonal basis:
\begin{eqnarray} \nonumber
s:\hspace{1.0cm} && \psi_{000}\\ \nonumber
p:\hspace{1.0cm} && \psi_{100},\;
                    \psi_{010},\;
                    \psi_{001}\\ \nonumber
d:\hspace{1.0cm} && \psi_{200} - \psi_{020},\;
                    \psi_{011},\;
                    2\psi_{002} - \psi_{200} - \psi_{020},\;
                    \psi_{101},\;
                    \psi_{110}\\ \nonumber
f:\hspace{1.0cm} && \psi_{300} - 3\psi_{120},\;
                    \psi_{201} - \psi_{021},\;
                   4\psi_{102} - \psi_{300} - \psi_{120},\;
                   \psi_{030} - 3\psi_{210}, \nonumber \\ 
                 &&   2\psi_{003} - 3\psi_{201} - 3\psi_{021},\;
                     4\psi_{012} - \psi_{210} - \psi_{030},\;
                    \psi_{111}. \label{eq140}
\end{eqnarray}
 The expansion of $\Psi_{\rm rel}$ now reads
 \begin{equation}
  \label{eq150}
  \Psi_{\rm rel}({\bf r},t) =
  \sum c_k(t) \, \psi_{\vec{l}_k}({\bf r},t)
 \end{equation}
 with the c-valued coefficients 
 \begin{equation}
  \label{eq160}
  c_k(t) = \langle \psi_{\vec{l}_k}({\bf r},t)|\Psi_{\rm rel}
  ({\bf r},t)\rangle
 \end{equation}
 and the index-vectors $\vec{l}_k = (l_k,m_k,n_k)$.
 The equation of motion for the expansion coefficients is 
 \begin{equation}
  \label{eq170}
  i\hbar\, {\bf \dot{c}} = \left(\tilde{H} - 
  \tilde{D}\right)\, {\bf c}
 \end{equation}
 with the time dependent coefficient vector ${\bf c}$, the
 Hamilton matrix 
 $\tilde{H}_{ij} = \langle \psi_{\vec{l}_i} | \hat{H} | 
 \psi_{\vec{l}_j} \rangle$, and the time evolution matrix
 $\tilde{D}_{ij} = \langle \psi_{\vec{l}_i} | i\hbar \partial_t | 
                   \psi_{\vec{l}_j} \rangle$.  
 
 Equation (\ref{eq170}) can be regarded as the quantum mechanical part
 of the time evolution, which is necessary to account for the deformation
 of the wavepacket. A second, classical contribution is given by the 
 motion of the wavepacket-centre. The corresponding equation of motion
 is defined as
 \begin{equation}
  \label{eq180}
  {\bf \dot{P}} = 
  - \nabla_{\bf R} \langle \Psi_{\rm rel}({\bf r}) | V({\bf r}) |
                 \Psi_{\rm rel}({\bf r}) \rangle
 \end{equation}
 with $V({\bf r}) = -e^2/|{\bf r}|$. The evaluation of the 
 matrix-elements is described in \cite{cle89}. The numerical solution
 of the equation of motion was done the same way as described in \cite{mer97},
 i.\,e.\ a second order integrator was used for the integration, the
 simulation was stopped after the Coulomb-potential has dropped to 2\%
 of its original value,
 and the final state was Fourier-transformed into momentum space. In order
 to evaluate the $2$-particle correlation function of the relative 
 momentum $q = |{\bf p}_1-{\bf p}_2|$,
 \begin{equation}
   \label{eq190}
   {\cal C}_{2}(q) = \frac{{\cal P}_{2}({\bf p}_1, {\bf
   p}_2)}{ {\cal P}({\bf p}_1)\, {\cal P}({\bf p}_2)}\;,
\end{equation}
a Monte-Carlo procedure \cite{mer96}
was used to sample the $2$-particle probability 
density ${\cal P}_{2}({\bf p}_1, {\bf p}_2)$, and event-mixing was
used to obtain the denominator of Eq.\ (\ref{eq190}).  
As described in \cite{mer97}, it is possible to generate a large number
$N_s$ of pairs from one single momentum state. After a number of $N_e$
independent events are calculated, the total number of pairs equals 
$N_p = N_e\cdot N_s$. 

\section{Results}
\label{results}
 As source function, i.\,e.\ the initial phase space distribution
 of the wavepackets, we applied the following simple parametrization: 
 \begin{eqnarray}
 \label{eq200}
  \rho ({\bf R})&=&(\pi R_s^2)^{-3/2}
  \exp\left( \frac{-{\bf R}^2}{R_s^2} \right)\;,\\
   f({\bf P}) &=& (2\pi m T)^{-3/2}\, \exp\left(\frac{-{\bf P}^{2}}{2 m
   T}\right)\;, \label{eq210}
 \end{eqnarray}
 i.\,e.\ the source was a static Gaussian with rms- radius $\sqrt{3/2}R_s$
 and temperature $T$. The initial width $\sigma_o$ of the wavepackets
 is to some extent arbitrary, but there exist certain upper and lower
 limits. Since any localization of a quantum state is accompanied with
 the corresponding zero point energy
 \begin{equation}
 \label{eq220}
  E_o = \frac{3\hbar^2}{8m\sigma_o^2}\;,
 \end{equation}
 the observed pion energies \cite{pel97a,pel97b}
 do not allow the localization to be well
 below $\approx 1$ fm, which already 
 yields $E_o(\sigma_o = 1 \mbox{fm}) \approx 100$ MeV.
 On the other hand, the correlation radii which are observed
 in correlations of identical pions also depend
 on $\sigma_o$, because the effective radius of the source
 is given in 1.\ order by
 \begin{equation}
 \label{eq225}
 R_{\rm eff} = \sqrt{R_s^2 + 2 \sigma_o^2}\;.
 \end{equation}
 Since the measured radii have a typical size 
 $R_{\rm corr} \approx 6$ fm \cite{bar97}, the upper limit
 for $\sigma_o$ may not significantly exceed $\approx 4$ fm. 

\begin{figure}
  \epsfxsize=16.0cm
  \epsffile{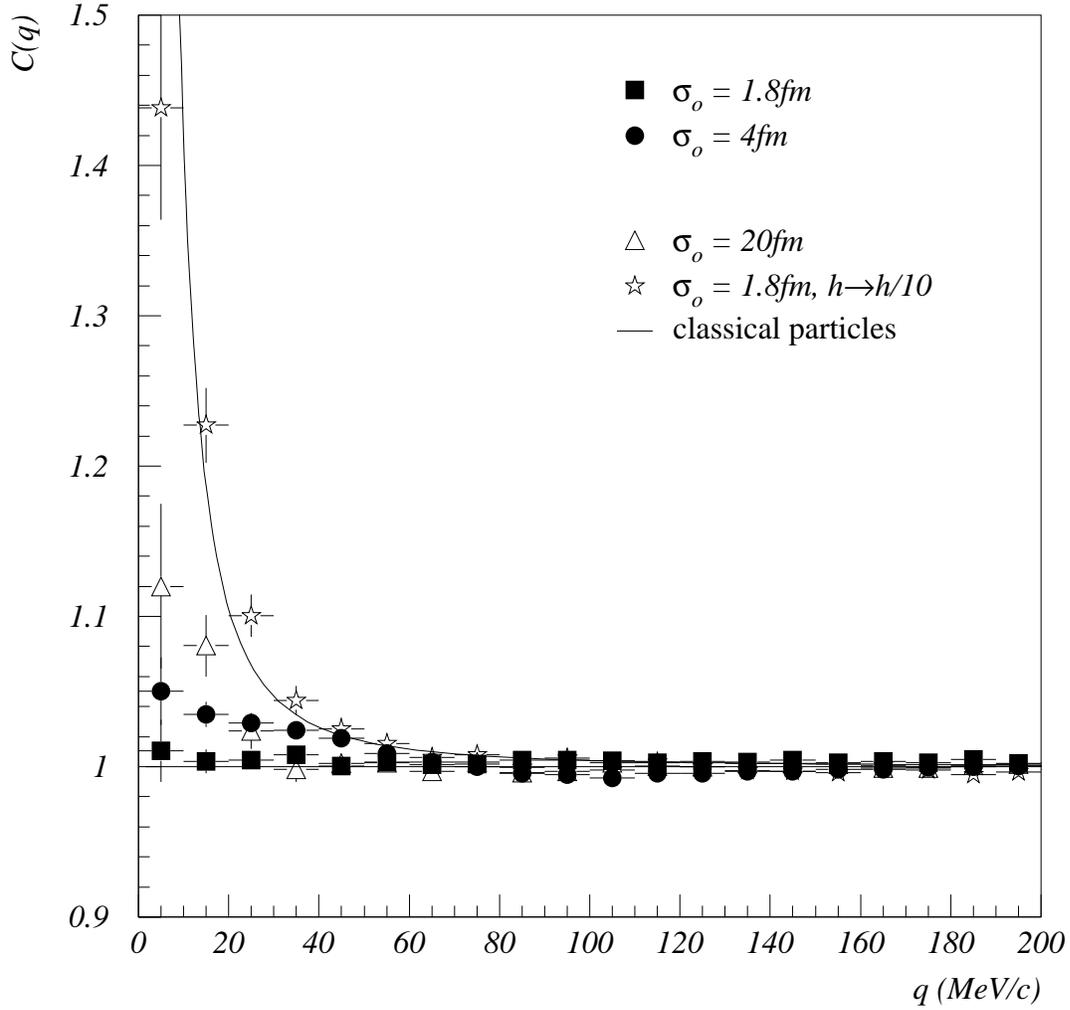}
  \caption{\label{fig1} 
  The correlation function for $\pi^+$--$\pi^-$ pairs with
  different initial localizations $\sigma_o$. Systems
  with $\sigma_o \le 4$ fm (solid symbols) are compatible with
  the results from like-charge pion interferometry. 
  The simulation with reduced Planck-constant (stars) displays the
  approach to the classical limit (solid curve). Number of generated
  pairs: $5\cdot 10^7$ (squares), $3\cdot 10^7$ (circles),
  $5\cdot 10^6$ (triangles), $5\cdot 10^6$ (stars).} 
 \end{figure}

 In Fig.\ \ref{fig1}  the correlation functions are shown for
 3 different initial wavepacket widths. In one of the 
 cases ($\sigma_o = 4$ fm, $R_s = 0$ fm) the width of the
 wavepacket determines the source size. In a second
 case, $\sigma_o = 1.8$ fm and $R_s = 5$ fm yield the
 same effective source size $R_{\rm eff} \approx 5.7$ fm
 as in the first case.
 The simulation with $\sigma_o = 20$ fm corresponds to the
 effective source size $R_{\rm eff} = 28$ fm, it is therefore 
 not compatible
 with the measured radii in like-charge pion interferometry
 and should be considered as a pure exercise to study the
 dependence of the Coulomb effect on $\sigma_o$.  
 In all cases the source temperature was set to $T = 50$ MeV.
 The states were evaluated
 using the $spd$-basis, in test simulations it was checked
 that the increase to the larger $spdf$-basis did not change
 the results in any observable way. The following facts are
 important:
 \begin{enumerate}
 \item The enhancement in the correlation function for small
  relative momenta $q$ is much weaker than obtained in the
  classical trajectory calculation (solid curve). 
 \item The enhancement becomes more pronounced with an
 increasing value for the initial width $\sigma_o$ of the states.  
 \end{enumerate}
 These results confirm the observations made in simulations of
 identical pions \cite{mer97}. There, the 
 factual disappearance of the Coulomb signal was explained by the 
 dominance of the zero point energy over the Coulomb energy.
 The maximum Coulomb energy in a system of 2 Gaussian states
 is reached when both completely overlap, yielding
 (without interference, i.\,e.\ for nonidentical particles)
 \begin{equation}
 \label{eq230}
 E_{\rm coul}^{\rm max} = \frac{Z^2e^2}{\sqrt{\pi}\sigma_o}\;.
 \end{equation}
 We define the ratio
  \begin{equation}
 \label{eq240}
 \xi \equiv \frac{E_{\rm coul}^{\rm max}}{E_o} =
 \frac{8 e^2}{3\sqrt{\pi} \hbar^2} \cdot Z^2 m \sigma_o
 \end{equation}
 as the {\it relative Coulomb strength} of the system. We
 obtain $\xi = 1.4\cdot 10^{-2}$ for the $\sigma_o = 1.8$ fm case,
 $\xi = 3.1\cdot 10^{-2}$ for $\sigma_o = 4$ fm 
 and $\xi = 0.16$ for $\sigma_o = 20$ fm. 
 Due to the smallness of $\xi$, in all cases
 the system is dominated by the quantum dispersion.
 As a crosscheck, the fundamental constant $\hbar$ was reduced
 to $1/10$ of its real value. This implies a reduction of
 $E_o$ by a factor of 100 and, for $\sigma_o = 1.8$ fm, a 
 relative Coulomb strength $\xi = 1.4$, which is now in the
 classical domain. The simulation results 
 (with source size $R_s = 5$ fm, stars) are
 already in good agreement with the corresponding classical 
 trajectory-calculation for pointlike particles
 ($R_s = 5.7$ fm, solid curve). 

 Similary, an increase in particle mass and/or charge
 can shift the ratio $\xi$ towards the classical domain. 
 For example, protons with $\sigma_o = 4$ fm correspond 
 to $\xi = 0.2$, $\alpha$-particles with $\sigma_o = 4$ fm 
 correspond to $\xi = 3.4$. In simulations, a strong 
 Coulomb signal was observed for these particles.
 Nevertheless, for small relative 
 momenta, a deviation from the classical results remains, 
 since the momentum uncertainty $\sigma_p = 
 \hbar/(2\sigma_o) = 25$ MeV/c does not depend
 on the particle mass, but reduces the momentum
 resolution which becomes significant for the rapid rise of
 the Coulomb signal when $q$ approaches zero.

 \section{Discussion}
 In proton- \cite{afa97} or nucleus- \cite{bar97} induced 
 reactions with energies above $10$ AGeV, 
 rather strong positive correlations
 in $\pi^+$--$\pi^-$ pairs have been observed. 
 These are interpreted as a
 Coulomb effect --- a conclusion which is
 supported by classical trajectory 
 and/or plane-wave scattering calculations. In this work and in 
 \cite{mer97}
 we have tried to argue that both assumptions are not applicable for
 particles which are emitted from a source, the size of which is
 2 orders of magnitude smaller than the Bohr-radius of $\pi^+$--$\pi^-$
 pairs. Because of Heisenberg's uncertainty principle a spatial
 uncertainty of size $\sigma_o$ transforms into a momentum uncertainty
 which allows the mutual particle interactions to become discernible
 in momentum correlations only when the parameter $\xi$ in Eq.\
 (\ref{eq240}) is of order 1. The value of $\xi$ can increase
 by either increasing the interaction strength, the delocalization
 of the states or the reduced system mass. For the $\pi^+$--$\pi^-$
 system, the corresponding parameters under normal conditions are too
 small. Therefore, it is highly desirable to
 find an interpretation of the $\pi^+$--$\pi^-$ signal.
 Obviously, there are two ways out of the dilemma:
 \begin{enumerate}
 \item The observed correlations {\it are} of Coulomb origin,
 and there is a yet unknown inconsistency in the wavepacket
 model and/or its interpretation. For example,
 if the pion-states could be initiated with an
  arbitrary large wavepacket width
 $\sigma_o$, then the ratio $\xi$ in Eq.\ (\ref{eq240}) easily
 could reach values above 1 and hence the classical (plane-wave) 
 assumption would be justified.
 \item The observed correlations have another origin,
 for example strong interactions, or residual dynamical
 correlations which could not be eliminated in the event
 mixing procedure during data analysis.
 \end{enumerate}
 With respect to the first point, i.\,e.\ the wavepacket 
 size, one has to bear in mind  
 that the pion wavefunction is interpreted as the
 probability amplitude to detect the pion at position ${\bf r}$
 in case of a measurement, and there is no physical reason
 to assume that the pion can be found far out of the source
 already at emission time. Additionally, the short lifetime 
 of the $\Delta(1232)$-resonance, the scattering and
 reabsorption by the surrounding nuclear matter and, particulary,
 the experimentally observed small values for
 the effective source sizes
 do not suggest the description of the pion state by 
 a widely extented coherent wavefunction.
 Consequently, $\sigma_o$ cannot
 exceed the size of the source, especially since that size
 is to be determined in the interferometry of like-charge pions.
 In order to measure an object with a given spatial extent $R_s$,
 one unavoidably has to cope with the corresponding momentum
 uncertainty. 

 But there exists a third and to some extent speculative 
 possibility, which has also been discussed in \cite{afa97}:
 With increasing energies, more and more long-living 
 resonances are formed, which do not decay within a small volume. 
 Pions which are produced under such circumstances
 do not necessarily fall under the above considerations,
 i.\,e.\ they may approximately be described as plane
 waves. These pions could contribute to the observed 
 $\pi^+$--$\pi^-$ peak, but this process would not work
 for $\pi^+$--$\pi^+$ and $\pi^-$--$\pi^-$ pairs emitted from the
 short lived and smaller interaction zone formed in heavy ion
 collisions.
 Consequently, the measured correlation function
 of like-charge pions should be corrected by a Coulomb factor
 which is weighted with the fraction of pions produced by
 long-living resonances. This hypothesis does not
 explain why the observed $\pi^+$--$\pi^-$ correlations are
 of the observed strength, since evidently not all pions
 could have been produced far outside the interaction zone
  --- otherwise,
 there wouldn't be any visible Bose-Einstein signal in the
 like-charge data. It seems that low energy data, which exclude
 the influence of long-living resonances, could help to 
 solve the $\pi^+$--$\pi^-$ puzzle.

\begin{thebibliography}{hol93}
\bibitem{hbt57} R. Hanbury Brown and R. Q. Twiss, Proc.\ Roy.\ Soc.\ 
                (London) A 243, 291 (1957).
\bibitem{gol60} G.\ Goldhaber et al.\,, Phys.\ Rev.\ 120, 300 (1960).
\bibitem{hol95} P.\,R.\ Holland, ``The Quantum Theory of Motion'',
                Cambridge University Press 1993.
\bibitem{mer96} H.\ Merlitz and D.\ Pelte, Z.\ Phys.\ A 357 (1997) 175.
\bibitem{cso97}  T.\ Cs\"org\"o and J.\ Zim\'anyi,
                Phys.\ Rev.\ Lett.\ 80 (1998) 916.
\bibitem{wie98} U.\,A.\ Wiedemann, nucl-th/9801009.  
\bibitem{bow91} M.\,G.\ Bowler, Phys.\ Lett.\ B 270 (1991) 69.
\bibitem{biy95} M.\ Biyajima, T.\ Mizoguchi, T.\ Osada and G.\ Wilk,
                Phys.\ Lett.\ B 353 (1995) 340.
\bibitem{sin96} Yu.\,M.\ Sinyukov, S.\,V.\ Akkelin and A.\,Yu.\ Tolstykh,
                Nucl.\ Phys.\ A 610 (1996) 278c.
\bibitem{bay96}  G.\ Baym and P.\ Braun-Munzinger, Nucl.\ Phys.\ A 610 
                (1996) 286c. 
\bibitem{mer97} H.\ Merlitz and D.\ Pelte, Phys.\ Lett.\ B 415
                (1997) 411.
\bibitem{bar97} J.\ Barrette and the E877 Collaboration, Phys.\ Rev.\ Lett.\
                78 (1997) 2916.
\bibitem{cle89} E. Clementi, ``Modern Techniques in Computational
                Chemistry: MOTECC-89'', ESCOM Science Publishers B.\,V.\
                (1989).
\bibitem{pel97a} D.\ Pelte and the FOPI collaboration, Z.\ Phys. A
                 357, 215 (1997).
\bibitem{pel97b} D.\ Pelte and the FOPI collaboration, Z.\ Phys. A
                 359, 55 (1997).
\bibitem{afa97}  L.\,G.\ Afanasyev et al.\,, Phys.\ At.\ Nuc.\ 60,
                 938 (1997).  
\end{thebibliography}
\end{document}